\documentclass{iaus}
\usepackage{graphicx}
\usepackage{natbib}
\bibpunct{(}{)}{;}{a}{}{,}

\title[3D MHD simulations of barred galaxies] %% give here short title %%
{3D MHD simulations of magnetic fields and radio polarization of barred galaxies}

\author[B. Kulesza-\.Zydzik et al.]   %% give here short author list %%
{B. Kulesza-\.Zydzik$^1$
, K. Kulpa-Dybe{\l}$^1$, 
K. Otmianowska-Mazur$^1$,\break  
G. Kowal $^{1,2}$  
\and M. Soida$^1$}

\affiliation{$^1$Astronomical Observatory, Jagiellonian
University, ul Orla 171, 30-244 Krak\'ow, Poland \break \\[\affilskip]
$^2$Department of Astronomy, University of Wisconsin,
475 North Charter Street, Madison, WI 53706, USA \break }

\pubyear{2009}
\volume{259}  %% insert here IAU Symposium No.
\pagerange{119--126}
\date{?? and in revised form ??}
\jname{Cosmic Magnetic Fields: From Planets,
to Stars and Galaxies}
\editors{A.C. Editor, B.D. Editor \& C.E. Editor, eds.}
\begin{document}

\maketitle

\begin{abstract}
We present results of three-dimensional, fully nonlinear MHD simulations of 
a large-scale ma\-gnetic field evolution in a barred galaxy. 
The model does not take into consideration the dynamo process.
We find that the obtained magnetic field configurations are highly similar to the  
observed maps of the polarized intensity  of barred galaxies, because
the modeled vectors form coherent structures along the bar and spiral arms.
Due to the dynamical influence of the bar the gas forms spiral waves which go radially outward. Each spiral
arm forms the magnetic arm which stays much longer in the disk, than the gaseous spiral structure.
Addi\-tionally the modeled total energy of magnetic field grows due to strong compression
and shear of non-axisymmetrical bar flows and diffe\-rential rotation, respectively.
\end{abstract}

\firstsection 
\section{Model description}
We investigate the evolution of barred galaxy solving the resistive set of MHD equations. We apply an isothermal equation of state. Our galaxy is composed of 
four components: the large and massive halo, the central 
bulge, rotating disc of stars and finally the bar. The rotation curve of the stellar disc we deriveded from the isochrone potential. The bar component is described by the second order Ferrers ellipsoid \citep{ferrers-1877} with semi-axes $a=4$~kpc, $b=2$~kpc, $c=2$~kpc. It is initiated into the galaxy gradually in time until it reaches its final mass $M_{bar}=10^{10}\mathrm{M_{\odot}}$. In order to conserve the total mass of the galaxy we reduce the bulge mass, so we have $M_{bar}(t)+M_b(t)=$~const. during the calculations. The  bar angular velocity $\Omega_{bar}$ is set to be $25$~km~s$^{-1}$~kpc$^{-1}$.
We assume that the initial magnetic field is azimuthal ($B_z=0$, $B_r=0$, $B_{\varphi}(z,r)$)
and its distribution strictly depends on the gas distribution
via the following condition:
$\alpha=p_{mag}/p_{gas}.$
The computational domain extends from $-10$~kpc to
$10$~kpc in the $x$~and $y$ direction, and from $-2.5$~kpc to $2.5$~kpc in the
$z$ direction. In all models we use the same value  of the resistivity coefficient $\eta=3\cdot10^{25}$~cm$^2$s$^{-1}$ and the resolution $n_x=n_y=256$, $n_z=65$. 
We perform simulations of magnetic field evolution with a constant isothermal sound of speed $c_s=5$~km/s and $\alpha=0.001$ ($B_{\varphi0}=0.1\mu$G).
\section{Results}
\begin{figure}\begin{center}%\resizebox{1.\hsize}{!}
 {\includegraphics[width=0.3\columnwidth]{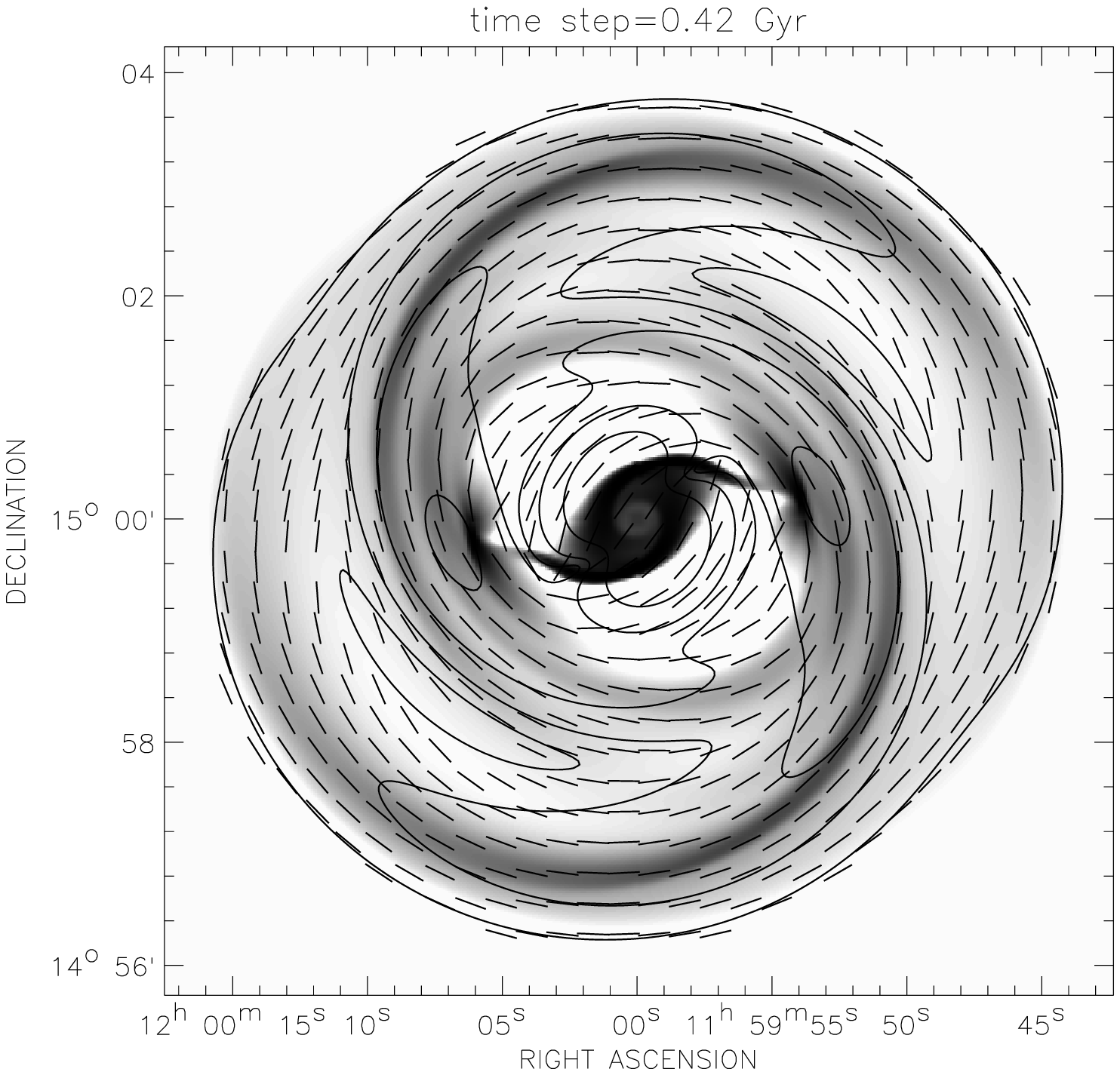}\includegraphics[width=0.3\columnwidth]{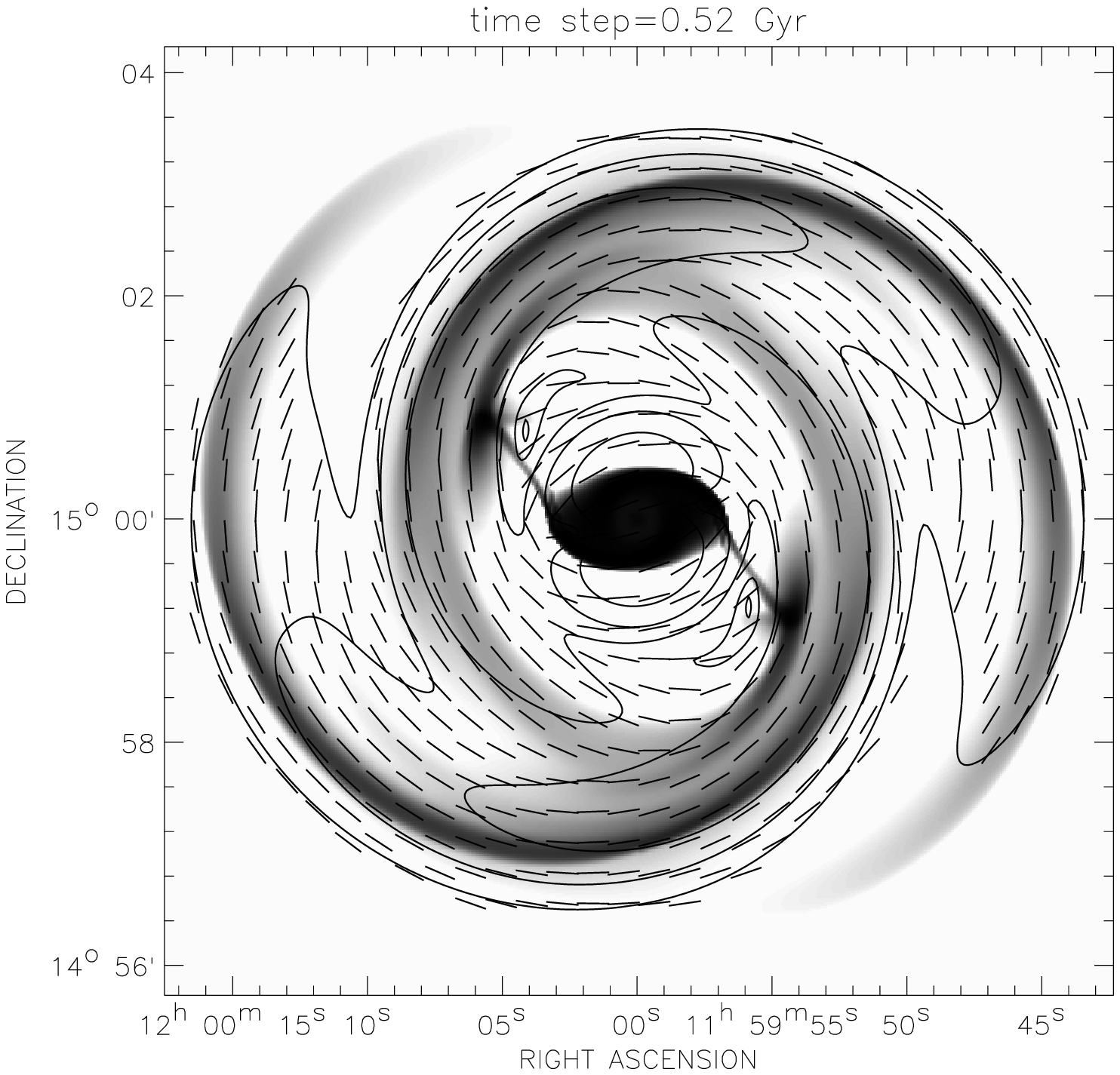}\includegraphics[width=0.3\columnwidth]{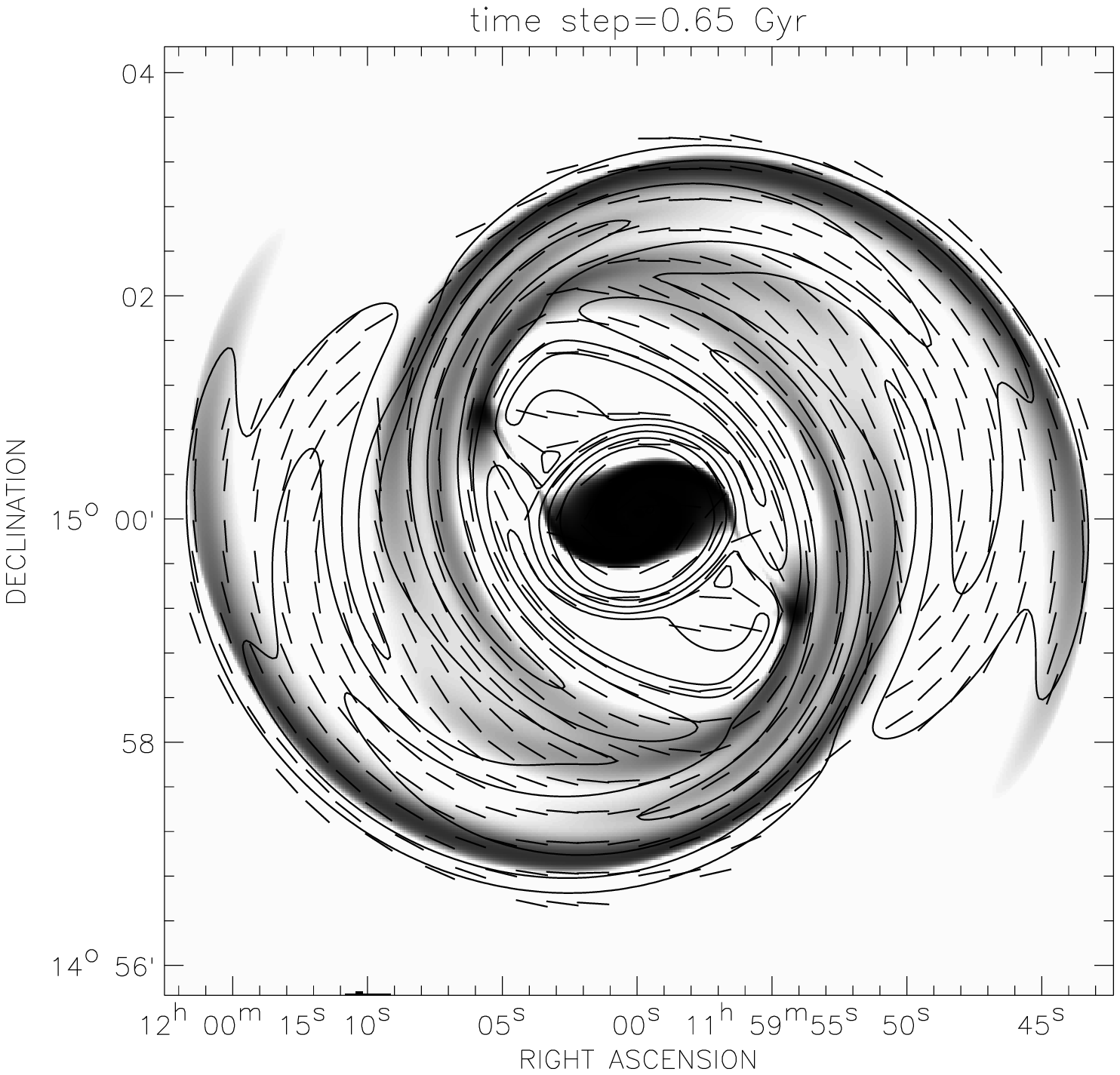}}
  \caption{Face-on polarization maps at $\lambda=6.2$~cm at selected times steps superimposed onto gaseous map. All maps have been smoothed to the resolution $25''$}
  \label{fig:polar}
\end{center}\end{figure}

Below we discuss the time evolution of the distributions of polarization angle and polarized intensity superimposed onto the column density. In order to show the magnetic field behaviour in our model
we present three crucial time steps (Fig.~\ref{fig:polar}):
\begin{itemize}
\item At time $t_1=0.42$~Gyr (Fig.~\ref{fig:polar}, left), as a consequence of the non-axisymmetric  gravitational potential of a bar, the gaseous and magnetic arms are formed. In the inner part of the disk, where the bar is presented, we obtain the highest density region and the strongest magnetic field. Moreover the magnetic field maxima are aligned along the gaseous ones.
\item As the simulation proceed, magnetic arms start to detach from gaseous spirals into the interarm regions (see $t_2=0.52$~Gyr Fig.~\ref{fig:polar}, in the middle).
\item At $t_3=0.65$~Gyr (Fig.~\ref{fig:polar}, right) our magnetic arms are also visible in the interam region. This is because the magnetic arms do not corotate with gaseous spirals but have a slightly lower angular velocity.
The process of drift of magnetic structures into the interarm area have also been obtained by \cite{otmian-02}.
\end{itemize}

We started our simulations with mean magnetic field equal $0.1$~$\mu$G. During the whole simulation time we observe the growth of the total magnetic energy. This amplification is caused by a local compression accompanied with leading sides of the bar and inner edges of spiral arms. As we do not apply the dynamo effect the mean value of the $B_{\phi}$ flux in the galactic midplane drops.

\section{Conclusions}
\begin{itemize} 
\item[1.]{We obtained the magnetic field vectors distributed along the bar, spiral
arms and also in the interarm region.}
\item[2.]{Magnetic arms are developing in the gaseous ones, but are detached
from the density waves. In the consequence the magnetic arms are shifted
to the interarm regions, what is in agreement with observations (e.g. NGC 1356 \citep{beck-05})} 
\item[3.]{The magnetic field energy in barred galaxies can be amplified without
any dynamo action but only due to non-axisymmetrical velocity.}
\end{itemize}
\begin{acknowledgments}
This work was supported by Polish Ministry of Science and Higher Education through grants: 92/N-ASTROSIM/2008/0, 2693/H03/2006/31 and 3033/B/H03/2008/35.
\end{acknowledgments}


\begin{thebibliography}{4}

\bibitem[{{Beck} {et~al.}(2005){Beck}, {Fletcher}, {Shukurov}, {Snodin}, {Sokoloff}, {Ehle}, {Moss} \& {Shoutenkov}}]{beck-05}
{Beck}, R., {Fletcher}, A., {Shukurov}, A., {Snodin}, A., {Sokoloff}, D. D., {Ehle}, M., {Moss}, D. \& {Shoutenkov}, V., 2005, A\&A, 444, 739

\bibitem[{{Ferrers}(1877){ferrers-1877}}]{ferrers-1877}  
 {Ferrers}, N. M., 1877, Quart. J. Pure Appl. Math., 14, 1

\bibitem[{{Otmianowska-Mazur} {et~al.}(2002){Otmianowska-Mazur}, {Elstner}, {Soida} \& {Urbanik}}]{otmian-02}
{Otmianowska-Mazur}, K., {Elstner}, D., {Soida}, M. \& {Urbanik}, M.,  2002, A\&A, 384, 48

\end{thebibliography}
\end{document}